\documentclass[preprint,12pt]{elsarticle}

\usepackage{amssymb}

\journal{XXX}

\begin{document}

\begin{frontmatter}



\title{Leveraging Interactions in Microfluidic Droplets for Enhanced Biotechnology Screens}


\author[puc]{Carolus Vitalis}
\author[puc]{Tobias Wenzel\corref{cor1}}

\affiliation[puc]{organization={Institute for Biological and Medical Engineering, Schools of Engineering, Medicine and Biological Sciences, Pontificia Universidad Católica de Chile},
            addressline={Vicuña Mackenna 4860}, 
            city={Macul},
            postcode={7820244}, 
            state={Santiago},
            country={Chile}}
\cortext[cor1]{corresponding author: tobias.wenzel@uc.cl}


\begin{abstract}
Microfluidic droplet screens serve as an innovative platform for high-throughput biotechnology, enabling significant advancements in discovery, product optimization, and analysis. This review sheds light on the emerging trend of interaction assays in microfluidic droplets, underscoring the unique suitability of droplets for these applications. Encompassing a diverse range of biological entities such as antibodies, enzymes, DNA, RNA, various microbial and mammalian cell types, drugs, and other molecules, these assays demonstrate their versatility and scope. Recent methodological breakthroughs have escalated these screens to novel scales of bioanalysis and biotechnological product design. Moreover, we highlight pioneering advancements that extend droplet-based screens into new domains: cargo delivery within human bodies, application of synthetic gene circuits in natural environments, 3D-printing, and the development of droplet structures responsive to environmental signals. The potential of this field is profound and only set to increase.
\end{abstract}




\end{frontmatter}



\pagebreak

\section{The Power of Interaction Screens in Droplets}
\label{sec:introduction}

In the rapidly advancing field of biotechnology, interaction screens in droplets are emerging as an increasingly versatile and powerful tool. 
Microfluidic droplets stand out for their ability to facilitate strong and rapid interactions~\cite{Sart2022CellCI, Tan2022TheEO}, and interaction assays are crucial for the discovery, optimization, and analysis of various biotechnology application fields, including 
metabolic engineering~\cite{Tauzin2020InvestigatingHI, Xu2020MicrofluidicAP, Bowman2019MicrodropletAssistedSO, Chijiiwa2019SinglecellGO},
synthetic biology genetic circuits~\cite{Prangemeier2020MicrofluidicPF, Gan2022HighThroughputRP, Koveal2022AHM, Tabuchi2022riboswitch},
the discovery of functional antibodies~\cite{Grard2020HighthroughputSA, Seah2017MicrofluidicST, Shembekar2018SingleCellDM},
enzymes~\cite{Stucki2021DropletMA, Holstein2021CellfreeDE},
and drugs such as antibiotics~\cite{Terekhov2018UltrahighthroughputFP, Kehe2019MassivelyPS, Mahler2021HighlyPD}.

Droplet screens are an ultra-high throughput technology that enables researchers to encapsulate interacting cells and reagents into thousands to tens of millions of microfluidic droplets with a pL to low-nL volume and efficiently analyze experimental results from each droplet, for example by 
fluorescence-based droplet sorting~\cite{Fu2021RecentAO, Xi2017ActiveDS}, DNA sequencing~\cite{Grard2020HighthroughputSA}, cultivation~\cite{Mahler2021HighlyPD}, or imaging~\cite{Zhang2020LinkedOA}.
Droplet-based screens, in general, offer several advantages over more traditional methods, such as their ultra-high throughput, reduced reagent and plastic consumable use, cost-effectiveness, and minimized contamination risk~\cite{Hu2021OneCA}, while improving the cultivability of microorganisms~\cite{Watterson2020DropletbasedHC, Hu2021OneCA, Yin2022ADM}, spheroids and organoids~\cite{Wang2023RecentMO, Sart2022CellCI}. 

In the context of automated screening methods, droplets combine the advantages of the high-throughput and single-cell resolution of flow cytometry with the versatility of liquid-handling robots to use non-cell-bound reagents and dyes in assays~\cite{Zeng2020HighThroughputST}. 
Despite these advantages, the implementation of droplet screens also requires development time investment, and should ideally be chosen when more traditional screens are not well suited for the biotechnologically desired process. 
There are, however, many notable droplet-based biotechnology screens that harness the above advantages but are not necessarily based on interactions, such as different types of 
single-cell analysis~\cite{Matua2019SingleCellAU, Jiang2023RecentAI} including quantification~\cite{Hou2022DropletbasedDP},
transcriptomics~\cite{Liu2019HighTG} and
genomics~\cite{Nishikawa2022ValidationOT, Pryszlak2021EnrichmentOG, Jing2022virology} e.g., to identify and characterize production strains,
and directed evolution in droplets~\cite{Stucki2021DropletMA}, which is often applied to enzyme and pathway optimization.

\begin{figure}
    \centering
    \includegraphics[width=0.75\textwidth]{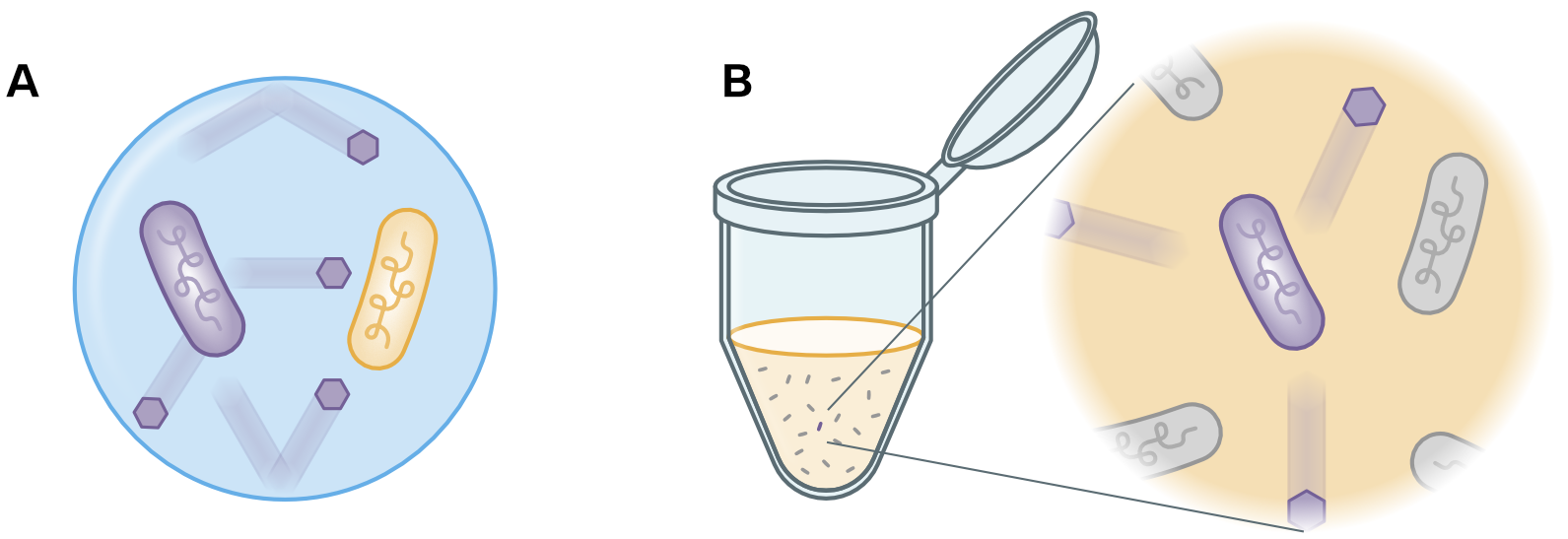}
    \caption{\textbf{Schematic representation of the advantage of droplets to facilitate interactions between cells.} \textbf{(A)} The compact volume of microfluidic droplets enables rapid elevation of concentrations of molecules secreted by cells. Simultaneously, it promotes the swift encounter of a few encapsulated molecules with their target. This fosters cellular and molecular interactions to take place rapidly and intensively, such as the survival of a dependent auxotrophic cell, symbolized here by a yellow cell. \textbf{(B)} In contrast, larger assay volumes in combinatorial screens dilute the limited cell numbers available, leading to low concentration levels. This could hinder cells from effectively communicating or exchanging metabolites within a viable time frame. For instance, auxotrophic strains might not survive (depicted by gray cells) in the population due to this delayed interaction.}
    \label{fig:dropletconcentration}
\end{figure}

Interaction assays are a unique strength of droplets, because they can usually not be performed with any other high-throughput method. 
Interactions in biotechnology are often based on combinatorial testing of a limited number of molecules and cells, which can achieve interactions in the confined space of microfluidic droplets~\cite{Tan2022TheEO}. 
Many such interactions would not take effect quickly enough in well-plates that contain microliters instead of picoliter volumes for the same number of molecules or cells to interact, and wells are also much more likely to contain contaminants in their larger volume. 
This volume-based advantage for interactions in droplets is illustrated in Figure\,\ref{fig:dropletconcentration}, based on a concrete example of metabolic interactions between auxotrophic bacterial strains by Tan et al.~\cite{Tan2022TheEO}.

This review paper is a succinct exploration of the recent literature, primarily from the last two years, that highlights diverse applications and the significant advancements in droplet interaction screens. The aim is not to provide an exhaustive account, but this article type rather highlights recent examples and trends that showcase the technique's versatility and potential in both microbial and mammalian cell screens in biotechnology. We will also highlight recent methodology advances that expand droplet-based interaction screens to new environments, including cargo delivery inside the body, application of synthetic gene circuits in the natural environment, 3D-printing, and more. It is the authors' belief that the interaction-focused perspective of droplet screens will lead to the discovery of previously unavailable screens that hold significant promise for scientific and commercial innovation.

\section{Recent Highlights in Droplet-Based Interaction Assays}
\label{sec:interactions}

\begin{figure}
    \centering
    \includegraphics[width=1\textwidth]{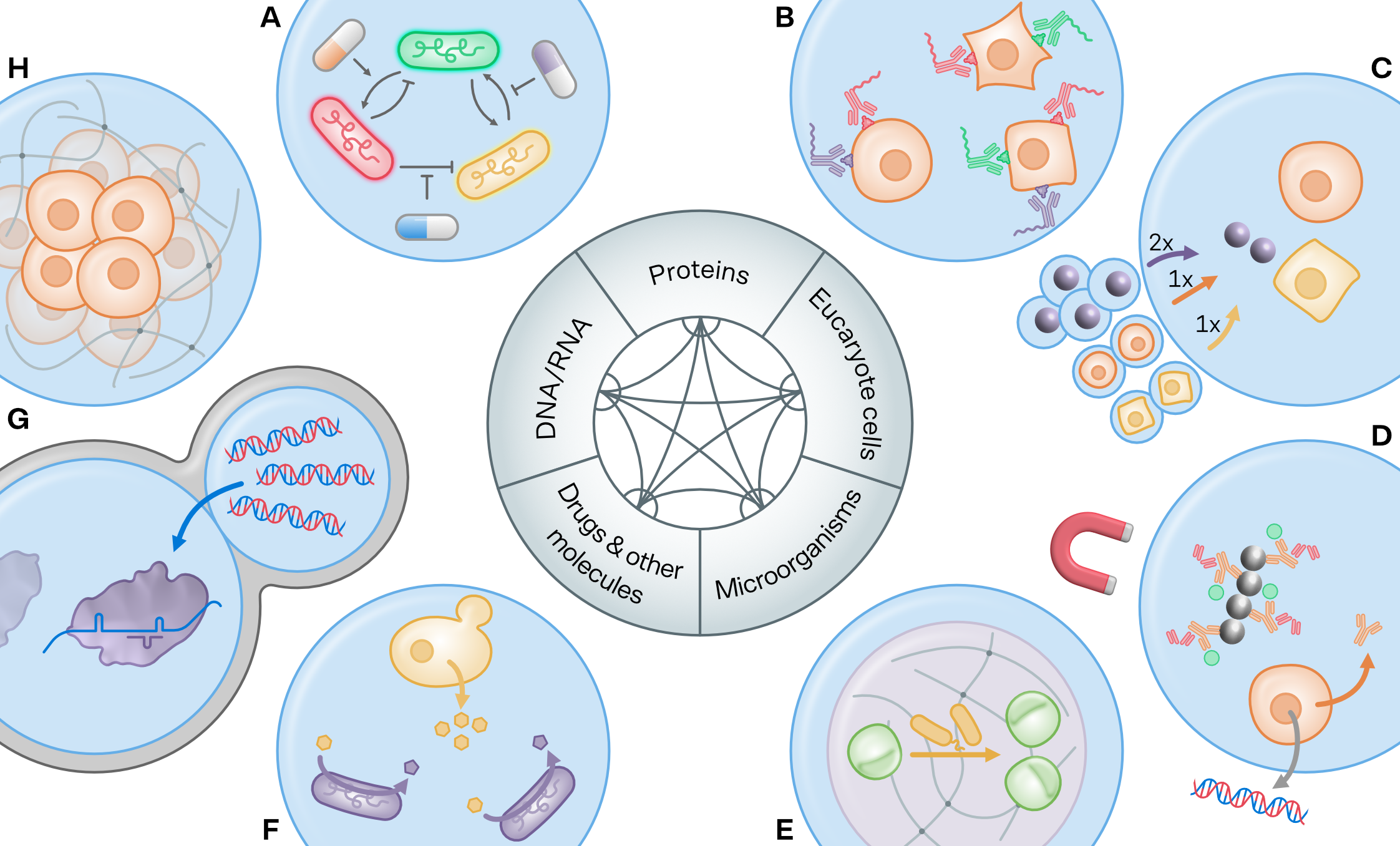}
    \caption{\textbf{Schematic representation of different kinds of droplet-based interaction screens.} The Interactions of many biological entities can be screened in droplets, including proteins such as antibodies and enzymes, eukaryotic cells, microorganisms, drugs, and other molecules, and DNA/RNA. At the center, binary combinatorial interaction possibilities are shown by connecting edges in a graph, and screens can incorporate more than two categories. Droplet schematics \textbf{A}-\textbf{H} highlight some examples of noteworthy interaction screens: (\textbf{A}) Quantitative network analysis of interactions between microorganisms and antibiotic drugs~\cite{Hsu2019MicrobialIN}; (\textbf{B}) Protein-protein interaction (co-occurrence) screens on the surface of different cell types~\cite{Hwang2021SCITOseqSC}; (\textbf{C}) Active merging of droplets for interaction screens where all droplets have the same content type combination of cells and beads~\cite{Madrigal2022CharacterizingCI}; (\textbf{D}) Co-localization screen for functional antibodies that bind antigens to magnetic beads in droplets. The functional screen is followed by transcriptomics analysis in barcoded droplets to sequence the functional antibody coding genes~\cite{Grard2020HighthroughputSA}; (\textbf{E}) Microbiota-algae co-cultivation screen in gel-microdroplets to select and sequence synergistic helper strains that aid the growth of algae culture~\cite{Ohan2019HighthroughputPO}; (\textbf{F}) Yeast production cell-biosensor interaction in droplets to quantitatively indicate excreted product concentrations~\cite{Siedler2017DevelopmentOA}; (\textbf{G}) DNA-protein interaction screen to detect all human viruses in multiple samples at once in a highly multiplexed diagnostic assay in a double-droplet trap array~\cite{Ackerman2020MassivelyMN}; (\textbf{H}) Cell-cell interactions to form spheroids and organoids in gel-microdroplets~\cite{Wang2023RecentMO}.}
    \label{fig:interactions}
\end{figure}

Droplet-based interaction assays have been successfully applied to study interactions between different cell types, proteins, DNA \& RNA, drugs, and other molecules, as illustrated in Figure\,\ref{fig:interactions}.
Literature examples can be found for all 15 interaction categories shown in the figure's central graph, and some methods combine more than two types of interacting entities.
While a wide spectrum of examples exists, most interaction screen types are still in the early stages of development, and we expect more research studies to be published in the coming years on this topic to mature biotechnological screening applications. 
Nevertheless, recent breakthroughs in droplet microfluidic screens for interaction assays have enabled new scales of bioanalysis and biotechnological product design. 
In the following paragraphs, we highlight recent examples of different types of droplet-based interaction screens and discuss their potential uses to demonstrate the power and versatility of this emerging method trend and stimulate its application in research and industry.

Microfluidic droplets serve as a powerful tool for the quantitative exploration of complex interaction networks among cells and supplementary substances like drugs across various concentrations. 
The microbial network analysis shown in Figure\,\ref{fig:interactions}\,A, for example, demonstrates an analysis involving three distinct, labeled auxotrophic bacterial cultures~\cite{Hsu2019MicrobialIN}. 
The potential for applications of such screens in modular genetic engineering approaches~\cite{Lu2019ModularME} is notable, as it offers the possibility to widen and enrich the biochemical synthesis space through identifying productive co-cultures~\cite{Arora2020ExpandingTC}. 
Scalable analysis of microbial metabolic interactions will further benefit the field of microbiome analysis and its biotechnology applications~\cite{Lawson2019CommonPA}.
It is also worth mentioning promising methods for microfluidic co-culture, which are not performed in droplets~\cite{Burmeister2019MicrofluidicCA}.

Protein-protein interactions constitute another major area where a variety of droplet methods have demonstrated their value, a topic thoroughly reviewed in another work~\cite{Arter2020MicrofluidicAF}. Figure\,\ref{fig:interactions}\,B illustrates a method where the co-occurrence of cell-surface proteins across different cell types is profiled utilizing a pool of labeled antibodies~\cite{Hwang2021SCITOseqSC}.

The encapsulation of interaction partners, like single cells or molecules, into droplets typically involves controlling the input concentration of each sample. This practice yields a stochastic encapsulation distribution in line with the Poisson distribution. As a result, achieving the preferred combination of different cells and molecules can be less efficient when dealing with larger quantities of interacting partners. To mitigate this challenge, several methods have been developed. One such strategy highlighted in Figure\,\ref{fig:interactions}\,C involves actively selecting and merging encapsulated entities into the desired combinations~\cite{Madrigal2022CharacterizingCI}. This approach has been employed in a screen to detect cytokines on beads released during the interaction of two different cell types.

Interaction screens offer the potential to encompass an array of interaction types. Consider, for instance, the functional antibody screen depicted in Figure\,\ref{fig:interactions}\,D. In this instance, droplets contain a cell expressing antibodies which, in turn, interact with an antigen and labeled antibodies, thereby co-locating fluorescence onto magnetic beads within the droplets. The screening process is continued by recovering functional cells and re-encapsulating them into droplets with reagents and DNA-barcoded gel-microdroplets that facilitate the recovery of the transcribed genes of the antibody heavy and light chain~\cite{Grard2020HighthroughputSA}. Beyond the bead-based assay, the authors also demonstrated a direct screen on target cells, akin to an earlier antibody gene-target cell screen conducted by a different research team~\cite{Shembekar2018SingleCellDM}. 

Numerous types of interaction screens can be instrumental in enhancing the production of biotechnological products, be it by metabolic engineering~\cite{Xu2020MicrofluidicAP, Bowman2019MicrodropletAssistedSO} or directed evolution~\cite{Stucki2021DropletMA}. 
A prime example is the potential for increased yield in algal culture when cells are co-cultivated syntrophically with certain microorganisms, as opposed to monocultures.
Figure\,\ref{fig:interactions}\,E features a screen that selects synergistic bacterial strains through the utilization of the autofluorescence of proliferating \textit{Chlorella sp.} colonies in gel-microdroplets~\cite{Ohan2019HighthroughputPO}. The study also indicated that co-cultures in droplets could be incubated for a duration of up to 60 days. 
An analogous co-culture strategy was employed to culture microorganisms in the laboratory that did not grow in conventional cultures~\cite{Tan2020CocultivationOM}. 

Where autofluorescence and fluorescent labels are not readily available, synthetic biology biosensors can be incorporated into droplet screens to quantify the productivity of target cells by interaction~\cite{Koveal2022AHM} for metabolic engineering applications. The method outlined in Figure\,\ref{fig:interactions}\,F leverages droplets for yeast metabolic engineering to quantify the product (p-coumaric acid) excretion into the droplet with biosensor reporter cells. These biosensor cells harbor an operon with a transcription factor that responds to the product and encodes for a fluorescent protein signal~\cite{Siedler2017DevelopmentOA}. In more recent developments, a cellular biosensor was designed to screen for the secretion of the industrial chemical (3-hydroxypropionic acid) in droplets~\cite{Kim2021SyntheticCC}, and another recent biosensor platform evolved the yield (of erythromycin) from genetically difficult to engineer actinomycetes production strains~\cite{hua2022biosensor}.

Droplets do not always need to be sorted in order to be characterized. As demonstrated by the diagnostic assay illustrated in Figure\,\ref{fig:interactions}\,G, it is possible to screen multiple clinical samples for all 169 human-associated viruses (i.e., viruses with $>$10 published sequences) in a protein-DNA interaction screen. Initially, all DNA-amplified patient sample-derived droplets and the Cas13-based virus detection mix droplets were color-barcoded. Subsequently, droplets were trapped pairwise side-by-side in a large double-droplet trap array. Finally, the color barcode combinations were imaged before and after the droplet merger, measuring the color barcodes and a fluorescent readout where the target viral sequences were present~\cite{Ackerman2020MassivelyMN}.

Another notable application of the DNA-protein interaction screen type in biotechnology deserves mentioning. 
This involves the interplay between genetic fragments and the enzymes and metabolites of cell-free extracts, which serves to express and evaluate synthetic biology constructs within droplets. This method can significantly enhance the efficiency of engineering test cycles~\cite{Gan2022HighThroughputRP, Tabuchi2022riboswitch}.

Mammalian cell-cell interactions are crucial for the formation of physiologically relevant tissue, a critical area for the enhancement of early-stage clinical trials. Droplets are suitable for the generation of both spheroids and organoids, frequently in gel-microdroplets, such as Matrigel, agarose, alginate, gelatin, among others (refer to Figure\,\ref{fig:interactions}\,H), or within gel-shell capsules~\cite{Wang2023RecentMO}.

\section{Expanding Applications of Droplet-Based Interactions Screens}
\label{sec:materials}

As we have seen, high-throughput interaction screens based on microfluidic droplets can be performed with a versatile range of droplet components, such as cells and molecules, but also magnetic beads, color or DNA barcodes, and gels. Notably, advancements in gel-microdroplet materials have paved the way for a broader application of droplets in cell cultures, multi-step procedures, and streamlined molecular processing protocols. In this section, we highlight several innovative material developments that promise to broaden the reach of droplet-based interaction screens, catering to a wider audience and novel environments, including interactions occurring between droplets and their surrounding milieu.

\begin{figure}
    \centering
    \includegraphics[width=1\textwidth]{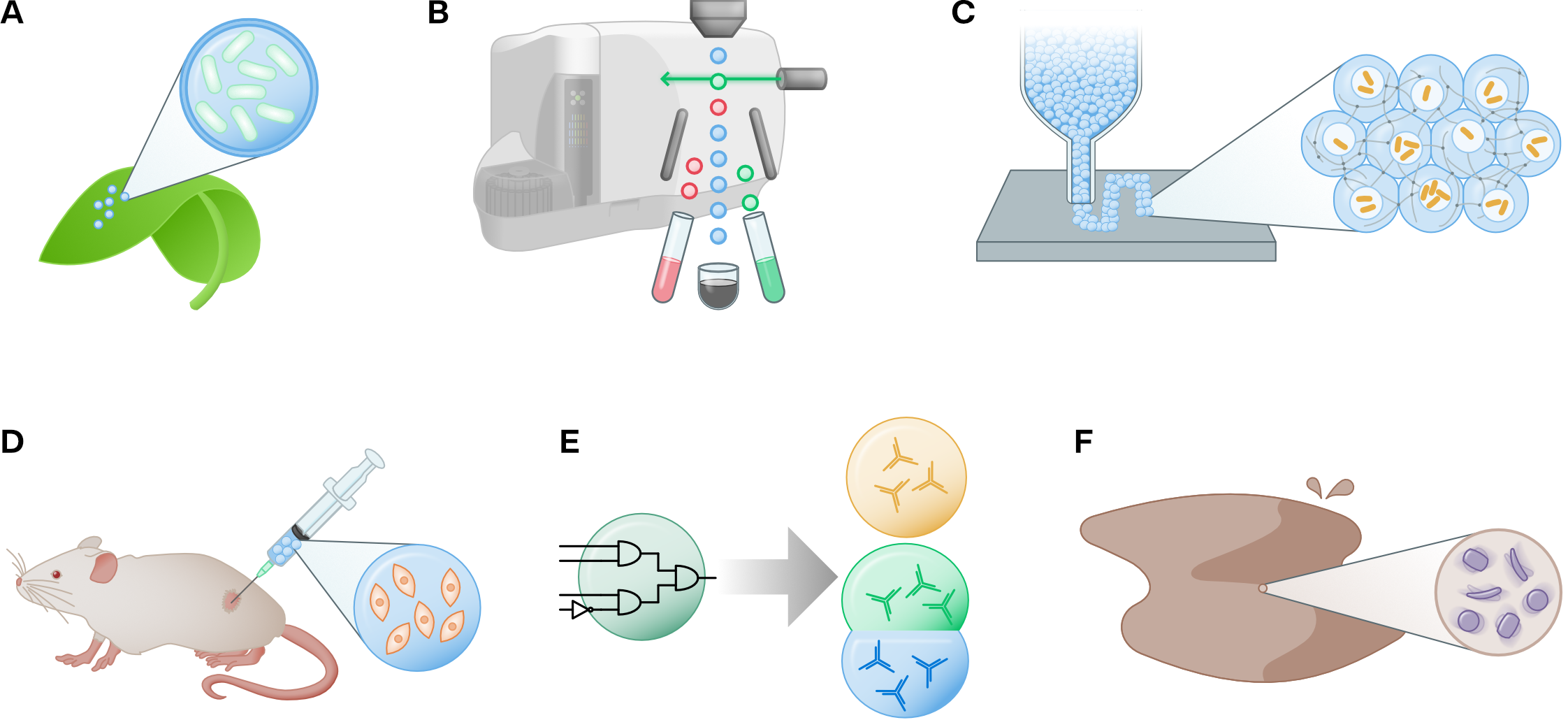}
    \caption{\textbf{Illustration of emerging droplet-based interaction screens facilitated by using new materials.} \textbf{(A)} Biocontainment of microorganisms is enabled by an alginate-polyacrylamide coating that allows controlled incorporation of these organisms into the environment~\cite{Tang2021HydrogelbasedBO}. \textbf{(B)} Cell sorting using a flow cytometer and double emulsion~\cite{Brower2020DoubleEF}. \textbf{(C)} Bioprinting of cells encapsulated in shell-droplets that deliver more precise spatial cell-cell interactions~\cite{Ou2022BioprintingMF}. \textbf{(D)} Hair follicles delivered by microfluidic droplets of gelatin methacryloyl and chitosan hydrogel, present a new method of targeted delivery~\cite{Zhang2022ScalableAH}. \textbf{(E)} DNA droplets contain a genetic circuit capable of sensing the presence of microRNAs by disrupting the homogeneous distribution of DNA, separating them into three distinct droplets~\cite{Gong2022ComputationalDD}. \textbf{(F)} Wastewater treatment using micromotors whose generation was assisted by microfluidic droplets and which catalyze the degradation of organic waste~\cite{Qu2022DropletMS}.}
    \label{fig:materials}
\end{figure}

For example, coated microgels (alginate hydrogels with an alginate and polyacrylamide coating) have been used to physically contain genetically modified microorganisms~\cite{Tang2021HydrogelbasedBO}, allowing them to be incorporated into the natural environment in a controlled manner, as illustrated in Figure~\ref{fig:materials}\,A. Even after 72 h, no microorganism escape was detected, but nutrients and signal molecules were successfully exchanged between the environment and adjacent capsules. Such capsules provide the basis for syntrophic interaction screens with the environment. 

A widely used advantage of gel-microdroplets and double emulsion droplets is that they can be processed in an aqueous environment and therefore sorted in a regular flow cytometer~\cite{Brower2020DoubleEF, Li2021DropletFC}. Figure\,\ref{fig:materials}\,B illustrates the use of double emulsion for cell sorting in a commercial flow cytometer. Such machines are available in a wider range of research laboratories and can therefore increase access to droplet-based interaction screens. Unfortunately, the generation of double emulsions and gel microdroplets is more challenging than that of regular droplets.

Cells can be enclosed in shell droplets, which can be 3D-printed and cross-linked into cell-containing structures~\cite{Ou2022BioprintingMF}, as shown in Figure\,\ref{fig:materials}\,C. This process opens new possibilities to place interaction partners in space for interactions between different encapsulated organisms and their environments. This technology may also enable the co-design of organs on a chip with a spatially and compositionally engineered microbiome.

A key environment for the placement and release of drugs and organisms is the human body. Indeed, droplets can be used for injection and targeted delivery~\cite{Zhao2021InjectableMH}, such as injecting gel-microdroplets loaded with hair formation cells and plasma with platelets into hairless tissue for regeneration~\cite{Zhang2022ScalableAH}. The method illustrated in Figure\,\ref{fig:materials}\,D utilizes specialized hydrogels (gelatin methacryloyl (GelMA) and chitosan) for targeted delivery. Delivery systems can be designed in several layers and with multiple functions, such as featuring a self-renewable hydration layer (based on liposomes) for delivery into joints~\cite{Lei2022InjectableHM}.

Finally, the droplets can be designed to actively respond to or actuate their environment. While these examples still tend to be in the proof-of-concept stage, they allow us to think about new types of interaction screens.
Figure\,\ref{fig:materials}\,E shows an example of a responsive droplet used for a bio-based assay, a single droplet made of a purpose-designed DNA gel with different DNA motifs and linkers to bind them. The linkers feature a base complementarity for four different target miRNAs. In the presence of target miRNAs, the linkers hybridize to them, which unbinds the motif and compartmentalizes the droplet into three single-motive droplets~\cite{Gong2022ComputationalDD}.
Another example of a responsive microfluidic droplet is the stimulated drug-content ejection of a multiphase Janus microparticle~\cite{Feng2022OnestepFO}. Droplet-based micromotors are not responsive, but are constantly actuated ~\cite{Lin2020HydrogelBasedJM}, which can include catalytic interaction properties used for organic waste degradation in wastewater treatment~\cite{Qu2022DropletMS} (Figure\,\ref{fig:materials}\,F).

\section{Conclusion}
\label{sec:conclusion}

Droplet microfluidics has firmly established itself as an unrivaled high-throughput approach for screening in the realm of biotechnological applications. The significance of droplet-based screens is surging in parallel with the growing arsenal of available methods and tools. A noteworthy trend emerging in this field are interaction assays. Given their small volume, droplets are perfectly suited for facilitating and quantifying interactions, as demonstrated by numerous high-impact studies recently. We've brought attention to various types of interaction methods, spanning protein-protein, cell-cell, DNA-protein, and drug-microorganism interactions, amongst a host of other studies blossoming in this rapidly expanding area.

Beyond droplet contents, the material composition, processing protocols, and novel strategies for introducing droplets into the environment are being explored, such as deploying them in more widely available instruments, within the human body for organism and drug delivery, or even in natural settings. These innovative methods unlock new avenues for interaction studies where the droplet content, or the responsive droplet material itself, can interact with the biological milieu. We anticipate this will catalyze the emergence of new biotechnological applications.

Despite the many benefits of droplet microfluidic screens, several challenges remain, including the need for experimental knowledge, specialized equipment, and interdisciplinary communication skills~\cite{Ortseifen2020MicrofluidicsFB}. Advanced droplet selection set-ups are particularly promising but not widely available or easily replicated~\cite{Sun2023screening}. Additionally, methods often need to be re-optimized and benchmarked, which require time investment. It is therefore particularly practical to focus method development efforts on screens that are unique for droplets such as interaction assays, which do not simply compete with other methods over a cost or efficiency margin.

As a final note, it is important to recognize the growing significance of alternative measurement methodologies in single-cell analysis, such as mass spectrometry, Raman spectroscopy, and impedance spectroscopy, among others. Our review does not cover these methods, because these techniques have yet to make their major impact within droplet-based interaction studies, nevertheless, their potential to enhance this field is clear by offering direct label free metabolomics, proteomics and electrochemical analysis of droplet contents. We anticipate that these complementary tools will increasingly intersect with droplet-based systems in the future, broadening our scope of high-throughput screening capabilities.

As we move forward, the potential of droplets for interaction screens continues to grow. This brief review hopes to stimulate further research and industrial use of these promising techniques, and we look forward to the discoveries and innovations that will undoubtedly continue to emerge from this exciting field.

\section{Acknowledgements}
\label{sec:acknowledgements}

This work is part of a funded project granted to T.W. from ANID FONDECYT Iniciación 11200666. The funder had no role in study design, data collection and analysis, decision to publish, or preparation of the manuscript.

\section{Credit Author Statement}
\label{sec:contributions}

Tobias Wenzel: Funding acquisition, Supervision, and Writing – review \& editing.
Tobias Wenzel and Carolus Vitalis: Conceptualization, Data curation, Visualization, Writing – original draft.

\appendix

 \bibliographystyle{elsarticle-num} 
 \bibliography{cas-refs}

\begin{thebibliography}{10}
\expandafter\ifx\csname url\endcsname\relax
  \def\url#1{\texttt{#1}}\fi
\expandafter\ifx\csname urlprefix\endcsname\relax\def\urlprefix{URL }\fi
\expandafter\ifx\csname href\endcsname\relax
  \def\href#1#2{#2} \def\path#1{#1}\fi

\bibitem{Sart2022CellCI}
S.~Sart, G.~Ronteix, S.~Jain, G.~Amselem, C.~N. Baroud, Cell culture in
  microfluidic droplets., Chemical reviews (2022).

\bibitem{Tan2022TheEO}
J.~Y. Tan, T.~E. Saleski, X.~N. Lin, The effect of droplet size on syntrophic
  dynamics in droplet-enabled microbial co-cultivation, PLoS ONE 17 (2022).

\bibitem{Tauzin2020InvestigatingHI}
A.~Tauzin, M.~R. Pereira, L.~D. van Vliet, P.-Y. Colin, {\'E}.~Laville,
  J.~Esque, S.~Laguerre, B.~Henrissat, N.~Terrapon, V.~Lombard, M.~Leclerc,
  J.~Dor{\'e}, F.~Hollfelder, G.~Potocki-Veronese, Investigating
  host-microbiome interactions by droplet based microfluidics, Microbiome 8
  (2020).

\bibitem{Xu2020MicrofluidicAP}
P.~Xu, C.~A. Modavi, B.~Demaree, F.~F. Twigg, B.~T. Liang, C.~Sun, W.~Zhang,
  A.~R. Abate, Microfluidic automated plasmid library enrichment for
  biosynthetic gene cluster discovery, Nucleic Acids Research 48 (2020) e48 --
  e48.

\bibitem{Bowman2019MicrodropletAssistedSO}
E.~K. Bowman, H.~S. Alper, Microdroplet-assisted screening of biomolecule
  production for metabolic engineering applications., Trends in biotechnology
  (2019).

\bibitem{Chijiiwa2019SinglecellGO}
R.~Chijiiwa, M.~Hosokawa, M.~Kogawa, Y.~Nishikawa, K.~Ide, C.~Sakanashi,
  K.~Takahashi, H.~Takeyama, Single-cell genomics of uncultured bacteria
  reveals dietary fiber responders in the mouse gut microbiota, Microbiome 8
  (2019).

\bibitem{Prangemeier2020MicrofluidicPF}
T.~Prangemeier, F.~Lehr, R.~M. Schoeman, H.~Koeppl, Microfluidic platforms for
  the dynamic characterisation of synthetic circuitry., Current opinion in
  biotechnology 63 (2020) 167--176.

\bibitem{Gan2022HighThroughputRP}
R.~Gan, M.~D. Cabezas, M.~Pan, H.~Zhang, G.~Hu, L.~G. Clark, M.~C. Jewett,
  R.~Nicol, High-throughput regulatory part prototyping and analysis by
  cell-free protein synthesis and droplet microfluidics., ACS synthetic biology
  (2022).

\bibitem{Koveal2022AHM}
D.~Koveal, P.~C. Rosen, D.~J. Meyer, C.~M. D{\'i}az-Garc{\'i}a, Y.~Wang,
  L.~Cai, P.~J. Chou, D.~A. Weitz, G.~Yellen, A high-throughput multiparameter
  screen for accelerated development and optimization of soluble genetically
  encoded fluorescent biosensors, Nature Communications 13 (2022).

\bibitem{Tabuchi2022riboswitch}
T.~Tabuchi, Y.~Yokobayashi, High-throughput screening of cell-free riboswitches
  by fluorescence-activated droplet sorting, Nucleic Acids Research 50~(6)
  (2022) 3535--3550.
\newblock \href {https://doi.org/10.1093/nar/gkac152}
  {\path{doi:10.1093/nar/gkac152}}.

\bibitem{Grard2020HighthroughputSA}
A.~G{\'e}rard, A.~Woolfe, G.~Mottet, M.~Reichen, C.~Castrillon, V.~Menrath,
  S.~Ellouze, A.~Poitou, R.~Doineau, L.~Brise{\~n}o-Roa,
  P.~Canales-Herrer{\'i}as, P.~Mary, G.~Rose, C.~Ortega, M.~Delinc{\'e},
  S.~Essono, B.~Jia, B.~Iannascoli, O.~R.-L. Goff, R.~Kumar, S.~N. Stewart,
  Y.~Pousse, B.~Shen, K.~Grosselin, B.~Saudemont, A.~Sautel-Caill{\'e},
  A.~Godina, S.~McNamara, K.~Eyer, G.~A. Millot, J.~Baudry, P.~England,
  C.~Nizak, A.~Jensen, A.~D. Griffiths, P.~Bruhns, C.~J.~H. Brenan,
  High-throughput single-cell activity-based screening and sequencing of
  antibodies using droplet microfluidics, Nature Biotechnology 38 (2020)
  715--721.

\bibitem{Seah2017MicrofluidicST}
Y.~F.~S. Seah, H.~Hu, C.~A. Merten, Microfluidic single-cell technology in
  immunology and antibody screening., Molecular aspects of medicine 59 (2017)
  47--61.

\bibitem{Shembekar2018SingleCellDM}
N.~Shembekar, H.~Hu, D.~Eustace, C.~A. Merten, Single-cell droplet microfluidic
  screening for antibodies specifically binding to target cells, Cell Reports
  22 (2018) 2206 -- 2215.

\bibitem{Stucki2021DropletMA}
A.~Stucki, J.~Vallapurackal, T.~R. Ward, P.~S. Dittrich, Droplet microfluidics
  and directed evolution of enzymes: An intertwined journey, Angewandte Chemie
  (International Ed. in English) 60 (2021) 24368 -- 24387.

\bibitem{Holstein2021CellfreeDE}
J.~M. Holstein, C.~Gylstorff, F.~Hollfelder, Cell-free directed evolution of a
  protease in microdroplets at ultrahigh throughput, ACS Synthetic Biology 10
  (2021) 252 -- 257.

\bibitem{Terekhov2018UltrahighthroughputFP}
S.~S. Terekhov, I.~V. Smirnov, M.~Malakhova, A.~E. Samoilov, A.~I. Manolov,
  A.~S. Nazarov, D.~V. Danilov, S.~Dubiley, I.~A. Osterman, M.~P. Rubtsova,
  E.~S. Kostryukova, R.~H. Ziganshin, M.~Kornienko, A.~A. Vanyushkina,
  O.~Bukato, E.~N. Ilina, V.~V. Vlasov, K.~V. Severinov, A.~G. Gabibov,
  S.~Altman, Ultrahigh-throughput functional profiling of microbiota
  communities, Proceedings of the National Academy of Sciences of the United
  States of America 115 (2018) 9551 -- 9556.

\bibitem{Kehe2019MassivelyPS}
J.~Kehe, A.~B. Kulesa, A.~Ortiz, C.~M. Ackerman, S.~G. Thakku, D.~Sellers,
  S.~Kuehn, J.~Gore, J.~Friedman, P.~C. Blainey, Massively parallel screening
  of synthetic microbial communities, Proceedings of the National Academy of
  Sciences 116 (2019) 12804 -- 12809.

\bibitem{Mahler2021HighlyPD}
L.~Mahler, S.~P. Niehs, K.~Martin, T.~Weber, K.~Scherlach, C.~Hertweck,
  M.~Roth, M.~A. Rosenbaum, Highly parallelized droplet cultivation and
  prioritization of antibiotic producers from natural microbial communities,
  eLife 10 (2021).

\bibitem{Fu2021RecentAO}
X.~Fu, Y.~Zhang, Q.~Xu, X.~Sun, F.~Meng, Recent advances on sorting methods of
  high-throughput droplet-based microfluidics in enzyme directed evolution,
  Frontiers in Chemistry 9 (2021).

\bibitem{Xi2017ActiveDS}
H.-D. Xi, H.~Zheng, W.~Guo, A.~M. Ga{\~n}{\'a}n-Calvo, Y.~Ai, C.-W. Tsao,
  J.~Zhou, W.~Li, Y.~Huang, N.-T. Nguyen, S.~H. Tan, Active droplet sorting in
  microfluidics: a review., Lab on a chip 17 5 (2017) 751--771.

\bibitem{Zhang2020LinkedOA}
J.~Q. Zhang, C.~A. Siltanen, L.~Liu, K.-C. Chang, Z.~J. Gartner, A.~R. Abate,
  Linked optical and gene expression profiling of single cells at
  high-throughput, Genome Biology 21 (2020).

\bibitem{Hu2021OneCA}
B.~Hu, P.~Xu, L.~Ma, D.~Chen, J.~Wang, X.~Dai, L.~Huang, W.~Du, One cell at a
  time: droplet-based microbial cultivation, screening and sequencing, Marine
  Life Science \& Technology (2021).

\bibitem{Watterson2020DropletbasedHC}
W.~J. Watterson, M.~Tanyeri, A.~R. Watson, C.~M. Cham, Y.~Shan, E.~B. Chang,
  A.~M. Eren, S.~Tay, Droplet-based high-throughput cultivation for accurate
  screening of antibiotic resistant gut microbes, eLife 9 (2020).

\bibitem{Yin2022ADM}
J.~Yin, X.~Chen, X.~Li, G.~Kang, P.~Wang, Y.~Song, U.~Z. Ijaz, H.~Yin,
  H.~Huang, A droplet-based microfluidic approach to isolating functional
  bacteria from gut microbiota, Frontiers in Cellular and Infection
  Microbiology 12 (2022).

\bibitem{Wang2023RecentMO}
Y.~Wang, M.~Liu, Y.~Zhang, H.~Liu, L.~Han, Recent methods of droplet
  microfluidics and their applications in spheroids and organoids., Lab on a
  chip (2023).

\bibitem{Zeng2020HighThroughputST}
W.~Zeng, L.~Guo, S.~Xu, J.~Chen, J.~Zhou, High-throughput screening technology
  in industrial biotechnology., Trends in biotechnology (2020).

\bibitem{Matua2019SingleCellAU}
K.~Matuła, F.~Rivello, W.~T.~S. Huck, Single‐cell analysis using droplet
  microfluidics, Advanced Biosystems 4 (2019).

\bibitem{Jiang2023RecentAI}
Z.~Jiang, H.~Shi, X.~Tang, J.~Qin, Recent advances in droplet microfluidics for
  single-cell analysis, TrAC Trends in Analytical Chemistry (2023).

\bibitem{Hou2022DropletbasedDP}
Y.~Hou, S.-P. Chen, Y.~Zheng, X.~Zheng, J.-M. Lin, Droplet-based digital pcr
  (ddpcr) and its applications, TrAC Trends in Analytical Chemistry (2022).

\bibitem{Liu2019HighTG}
L.~Liu, C.~K. Dalal, B.~M. Heineike, A.~R. Abate, High throughput gene
  expression profiling of yeast colonies with microgel-culture drop-seq, Lab
  Chip 19 (2019) 1838--1849.

\bibitem{Nishikawa2022ValidationOT}
Y.~Nishikawa, M.~Kogawa, M.~Hosokawa, R.~Wagatsuma, K.~Mineta, K.~Takahashi,
  K.~Ide, K.~Yura, H.~Behzad, T.~Gojobori, H.~Takeyama, Validation of the
  application of gel beads-based single-cell genome sequencing platform to soil
  and seawater, ISME Communications 2 (2022).

\bibitem{Pryszlak2021EnrichmentOG}
A.~Pryszlak, T.~Wenzel, K.~W. Seitz, F.~Hildebrand, E.~Kartal, M.~R. Cosenza,
  V.~Benes, P.~Bork, C.~A. Merten, Enrichment of gut microbiome strains for
  cultivation-free genome sequencing using droplet microfluidics, Cell Reports
  Methods 2 (2021).

\bibitem{Jing2022virology}
W.~Jing, H.-S. Han, Droplet microfluidics for high-resolution virology,
  Analytical Chemistry 94~(23) (2022) 8085--8100.
\newblock \href {https://doi.org/10.1021/acs.analchem.2c00615}
  {\path{doi:10.1021/acs.analchem.2c00615}}.

\bibitem{Hsu2019MicrobialIN}
R.~H. Hsu, R.~L. Clark, J.~W. Tan, J.~C. Ahn, S.~Gupta, P.~A. Romero, O.~S.
  Venturelli, Microbial interaction network inference in microfluidic
  droplets., Cell systems (2019).

\bibitem{Hwang2021SCITOseqSC}
B.~Hwang, D.~S. Lee, W.~Tamaki, Y.~Sun, A.~Ogorodnikov, G.~C. Hartoularos,
  A.~Winters, B.~Z. Yeung, K.~L. Nazor, Y.~S. Song, E.~D. Chow, M.~H. Spitzer,
  C.~J. Ye, Scito-seq: single-cell combinatorial indexed cytometry sequencing.,
  Nature methods 18 8 (2021) 903--911.

\bibitem{Madrigal2022CharacterizingCI}
J.~L. Madrigal, N.~G. Schoepp, L.~Xu, C.~S. Powell, C.~L. Delley, C.~A.
  Siltanen, J.~Danao, M.~Srinivasan, R.~H. Cole, A.~R. Abate, Characterizing
  cell interactions at scale with made-to-order droplet ensembles (modes),
  Proceedings of the National Academy of Sciences of the United States of
  America 119 (2022).

\bibitem{Ohan2019HighthroughputPO}
J.~Ohan, B.~Pelle, P.~Nath, J.~Huang, B.~T. Hovde, M.~Vuyisich, A.~E.~K.
  Dichosa, S.~R. Starkenburg, High-throughput phenotyping of cell-to-cell
  interactions in gel microdroplet pico-cultures., BioTechniques 66 5 (2019)
  218--224.

\bibitem{Siedler2017DevelopmentOA}
S.~Siedler, N.~K. Khatri, A.~Zsoh{\'a}r, I.~Kj{\ae}rb{\o}lling, M.~D. Vogt,
  P.~Hammar, C.~F. Nielsen, J.~Marienhagen, M.~O.~A. Sommer, H.~N. Joensson,
  Development of a bacterial biosensor for rapid screening of yeast p-coumaric
  acid production., ACS synthetic biology 6 10 (2017) 1860--1869.

\bibitem{Ackerman2020MassivelyMN}
C.~M. Ackerman, C.~Myhrvold, S.~G. Thakku, C.~A. Freije, H.~C. Metsky, D.~K.
  Yang, S.~H. Ye, C.~K. Boehm, T.-S.~F. Kosoko-Thoroddsen, J.~Kehe, T.~G.
  Nguyen, A.~Carter, A.~B. Kulesa, J.~R. Barnes, V.~G. Dugan, D.~T. Hung, P.~C.
  Blainey, P.~C. Sabeti, Massively multiplexed nucleic acid detection with
  cas13, Nature 582 (2020) 277 -- 282.

\bibitem{Lu2019ModularME}
H.~Lu, J.~C. Villada, P.~K.~H. Lee, Modular metabolic engineering for biobased
  chemical production., Trends in biotechnology 37 2 (2019) 152--166.

\bibitem{Arora2020ExpandingTC}
D.~Arora, P.~Gupta, S.~Jaglan, C.~Roullier, O.~Grovel, S.~Bertrand, Expanding
  the chemical diversity through microorganisms co-culture: Current status and
  outlook., Biotechnology advances (2020) 107521.

\bibitem{Lawson2019CommonPA}
C.~E. Lawson, W.~R. Harcombe, R.~Hatzenpichler, S.~R. Lindemann, F.~E.
  L{\"o}ffler, M.~A. O’Malley, H.~G. Mart{\'i}n, B.~F. Pfleger, L.~Raskin,
  O.~S. Venturelli, D.~G. Weissbrodt, D.~R. Noguera, K.~D. McMahon, Common
  principles and best practices for engineering microbiomes, Nature Reviews
  Microbiology 17 (2019) 725--741.

\bibitem{Burmeister2019MicrofluidicCA}
A.~Burmeister, A.~Gr{\"u}nberger, Microfluidic cultivation and analysis tools
  for interaction studies of microbial co-cultures., Current opinion in
  biotechnology 62 (2019) 106--115.

\bibitem{Arter2020MicrofluidicAF}
W.~E. Arter, A.~Levin, G.~Krainer, T.~P.~J. Knowles, Microfluidic approaches
  for the analysis of protein–protein interactions in solution, Biophysical
  Reviews 12 (2020) 575 -- 585.

\bibitem{Tan2020CocultivationOM}
J.~Y. Tan, S.~Wang, G.~J. Dick, V.~B. Young, D.~H. Sherman, M.~A. Burns, X.~N.
  Lin, Co-cultivation of microbial sub-communities in microfluidic droplets
  facilitates high-resolution genomic dissection of microbial 'dark matter'.,
  Integrative biology : quantitative biosciences from nano to macro (2020).

\bibitem{Kim2021SyntheticCC}
S.~Kim, S.~H. Jin, H.~G. Lim, B.~Lee, J.~Kim, J.~Yang, S.~W. Seo, C.-S. Lee,
  G.~Y. Jung, Synthetic cellular communication-based screening for strains with
  improved 3-hydroxypropionic acid secretion., Lab on a chip (2021).

\bibitem{hua2022biosensor}
E.~Hua, Y.~Zhang, K.~Yun, W.~Pan, Y.~Liu, S.~Li, Y.~Wang, R.~Tu, M.~Wang,
  Whole-cell biosensor and producer co-cultivation-based microfludic platform
  for screening saccharopolyspora erythraea with hyper erythromycin production,
  ACS Synthetic Biology 11~(8) (2022) 2697--2708.
\newblock \href {https://doi.org/10.1021/acssynbio.2c00102}
  {\path{doi:10.1021/acssynbio.2c00102}}.

\bibitem{Tang2021HydrogelbasedBO}
T.-C. Tang, E.~Tham, X.~Liu, K.~Yehl, A.~J. Rovner, H.~Yuk, C.~de~la
  Fuente-Nunez, F.~J. Isaacs, X.~Zhao, T.~K. Lu, Hydrogel-based biocontainment
  of bacteria for continuous sensing and computation, Nature Chemical Biology
  17 (2021) 724 -- 731.

\bibitem{Brower2020DoubleEF}
K.~K. Brower, C.~Carswell-Crumpton, S.~L. Klemm, B.~Cruz, G.~Kim, S.~G.~K.
  Calhoun, L.~Nichols, P.~M. Fordyce, Double emulsion flow cytometry with
  high-throughput single droplet isolation and nucleic acid recovery, Lab on a
  chip 20 (2020) 2062 -- 2074.

\bibitem{Ou2022BioprintingMF}
Y.~Ou, S.~Cao, Y.~Zhang, H.~Zhu, C.~Guo, W.~Yan, F.~Xin, W.~Dong, Y.~Zhang,
  M.~Narita, Z.~Yu, T.~P.~J. Knowles, Bioprinting microporous functional living
  materials from protein-based core-shell microgels, Nature Communications 14
  (2023).

\bibitem{Zhang2022ScalableAH}
Y.~Zhang, P.~Yin, J.~Huang, L.~Yang, Z.~Liu, D.~Fu, Z.~Hu, W.~Huang, Y.~Miao,
  Scalable and high-throughput production of an injectable platelet-rich plasma
  (prp)/cell-laden microcarrier/hydrogel composite system for hair follicle
  tissue engineering, Journal of Nanobiotechnology 20 (2022).

\bibitem{Gong2022ComputationalDD}
J.~M. Gong, N.~Tsumura, Y.~Sato, M.~Takinoue, Computational dna droplets
  recognizing mirna sequence inputs based on liquid–liquid phase separation,
  Advanced Functional Materials 32 (2022).

\bibitem{Qu2022DropletMS}
C.~Qu, M.~Ren, Z.~Qiao, X.~Ren, W.~Guo, Droplet microfluidic synthesis of
  shape-tunable self-propelled catalytic micromotors and their application to
  water treatment, Journal of Materials Science 57 (2022) 20558 -- 20566.

\bibitem{Li2021DropletFC}
M.~Li, H.~Liu, S.~Zhuang, K.~Goda, Droplet flow cytometry for single-cell
  analysis, RSC Advances 11 (2021) 20944 -- 20960.

\bibitem{Zhao2021InjectableMH}
Z.~Zhao, Z.~Wang, G.~Li, Z.~Cai, J.~Wu, L.~Wang, L.~Deng, M.~Cai, W.~Cui,
  Injectable microfluidic hydrogel microspheres for cell and drug delivery,
  Advanced Functional Materials 31 (2021).

\bibitem{Lei2022InjectableHM}
Y.~Lei, Y.~Wang, J.~Shen, Z.~Cai, C.~Zhao, H.~Chen, X.~Luo, N.~Hu, W.~Cui,
  W.~Huang, Injectable hydrogel microspheres with self-renewable hydration
  layers alleviate osteoarthritis, Science Advances 8 (2022).

\bibitem{Feng2022OnestepFO}
Z.~Feng, B.~Zhou, X.~Su, T.~Wang, S.~L. Guo, H.~Yang, X.~Sun, One-step
  fabrication of multiphasic janus microparticles with programmed degradation
  properties based on a microfluidic chip, Materials \& Design (2022).

\bibitem{Lin2020HydrogelBasedJM}
X.~Lin, H.~Zhu, Z.~Zhao, C.~You, Y.~Kong, Y.~Zhao, J.~Liu, H.~Chen, X.~Shi,
  D.~Makarov, Y.~Mei, Hydrogel‐based janus micromotors capped with functional
  nanoparticles for environmental applications, Advanced Materials Technologies
  5 (2020).

\bibitem{Ortseifen2020MicrofluidicsFB}
V.~Ortseifen, M.~Viefhues, L.~Wobbe, A.~Gr{\"u}nberger, Microfluidics for
  biotechnology: Bridging gaps to foster microfluidic applications, Frontiers
  in Bioengineering and Biotechnology 8 (2020).

\bibitem{Sun2023screening}
G.~Sun, L.~Qu, F.~Azi, Y.~Liu, J.~Li, X.~Lv, G.~Du, J.~Chen, C.-H. Chen,
  L.~Liu, Recent progress in high-throughput droplet screening and sorting for
  bioanalysis, Biosensors and Bioelectronics 225 (2023) 115107.
\newblock \href {https://doi.org/10.1016/j.bios.2023.115107}
  {\path{doi:10.1016/j.bios.2023.115107}}.

\end{thebibliography}






\break\noindent\textbf{Special Interest Works}

\begin{itemize}

\item[{•}] R. Gan, M. D. Cabezas, M. Pan, H. Zhang, G. Hu, L. G. Clark, M. C. Jewett, R. Nicol, High-throughput regulatory part prototyping and analysis by cell-free protein synthesis and droplet microfluidics, ACS synthetic biology (2022).\\
This work shows an optimization screen for genetic circuits. It presents a high-throughput screening platform employing microfluidics, next-generation sequencing, and cell-free protein synthesis (CFPS). The study is a great example of how to use cell-free extract in droplets for versatile screens, especially in synthetic biology.

\item[{•}] Y. Wang, M. Liu, Y. Zhang, H. Liu, L. Han, Recent methods of droplet microfluidics and their applications in spheroids and organoids, Lab on a chip (2023).\\
Extensive and current overview of droplet microfluidic methods to form and trap different spheroids and organoids and a primer to their subsequent analysis. The article is helpful for researchers wanting to start in droplet-based organoids, as well as to readers wanting to advance using recent methods.

\item[{•}] A. Pryszlak, T. Wenzel, K. W. Seitz, F. Hildebrand, E. Kartal, M. R. Cosenza, V. Benes, P. Bork, C. A. Merten, Enrichment of gut microbiome strains for cultivation-free genome sequencing using droplet microfluidics, Cell Reports Methods 2 (2021).\\
The study presents a method to enrich and sequence microbiota from a complex gut microbiome sample in a targeted fashion. It uses droplet sorting and the interaction of microbial DNA with molecular probes to select and enrich target species in order to improve DNA sequencing quality. The study is an example of recent method innovation in precision genomics, and helps make the microbiome more experimentally accessible to biotechnology applications.

\item[{•}] J. Y. Tan, T. E. Saleski, X. N. Lin, The effect of droplet size on syntrophic dynamics in droplet-enabled microbial co-cultivation, PLoS ONE 17 (2022).\\
Systematic study demonstrating the effect of droplet size on metabolomic interactions between cells co-cultured in microfluidic droplets. It clearly and visually demonstrates the utility that especially small droplets have for effective and fast interaction-based screens. 

\item[{•}] J. L. Madrigal, N. G. Schoepp, L. Xu, C. S. Powell, C. L. Delley, C. A. Siltanen, J. Danao, M. Srinivasan, R. H. Cole, A. R. Abate, Characterizing cell interactions at scale with made-to-order droplet ensembles (modes), Proceedings of the National Academy of Sciences of the United States of America 119 (2022).\\
This work shows a method that allows a deliberate distribution of cells and molecules inside the microfluidic droplets of a screen. For this purpose, they combined two types of cells and beads (1:1:2) in a deterministic manner using a new combination of active droplet sorting and merging.

\item[{•}] T.-C. Tang, E. Tham, X. Liu, K. Yehl, A. J. Rovner, H. Yuk, C. de la Fuente-Nunez, F. J. Isaacs, X. Zhao, T. K. Lu, Hydrogel-based biocontainment of bacteria for continuous sensing and computation, Nature Chemical Biology 17 (2021) 724 – 731.\\
This work presents a hydrogel shell that allows physical containment. The contained cells could still perform their functions and interact with the environment but not physically leave the shell confinement. The method opens the possibility of incorporating synthetic genetic circuits in real-world scenarios.

\item[{•}] Y. Ou, S. Cao, Y. Zhang, H. Zhu, C. Guo, W. Yan, F. Xin, W. Dong, Y. Zhang, M. Narita, Z. Yu, T. P. J. Knowles, Bioprinting microporous functional living materials from protein-based core-shell microgels, Nature Communications 14 (2023).\\
Very recent article that shows how to combine cell encapsulation in microfluidic core-shell droplets with 3D printing. The study describes a method to cross-link the extruded droplet emulsion and demonstrates its application.

\item[{•}] Y. Zhang, P. Yin, J. Huang, L. Yang, Z. Liu, D. Fu, Z. Hu, W. Huang, Y. Miao, Scalable and high-throughput production of an injectable platelet-rich plasma (prp)/cell-laden microcarrier/hydrogel composite system for hair follicle tissue engineering, Journal of Nanobiotechnology 20 (2022).\\
Study presenting a new strategy for tissue engineering. It describes a new method of hair follicle delivery using microfluidic droplets, allowing the delivery of dermal papilla cells with platelet-rich plasma.

\end{itemize}

\end{document}